\documentclass[conference,a4paper]{IEEEtran}
\usepackage[T5,T1]{fontenc}
\usepackage[utf8]{inputenc}
\usepackage{booktabs}
\usepackage{array}
\usepackage{amsmath}
\usepackage{amssymb}
\usepackage{xcolor}
\usepackage{pgfplots}
\pgfplotsset{compat=1.18}
\usetikzlibrary{pgfplots.groupplots}
\usetikzlibrary{positioning,arrows.meta,fit}
\usepackage{url}
\usepackage[hidelinks]{hyperref}
\usepackage{orcidlink}
\hypersetup{pdftitle={REGATTA: A Cross-Domain Measurement Framework and Fault-Injection Harness for Enterprise AI Agents},pdfauthor={Luong Tuan Thanh and Abhijit Sanyal}}
\IEEEoverridecommandlockouts
\makeatletter
\def\@IEEEpubidpullup{4\baselineskip}
\makeatother
\IEEEpubid{\parbox{\textwidth}{\footnotesize
\textcopyright~2026 IEEE. Personal use of this material is permitted.
Permission from IEEE must be obtained for all other uses, in any current or
future media, including reprinting/republishing this material for advertising
or promotional purposes, creating new collective works, for resale or
redistribution to servers or lists, or reuse of any copyrighted component of
this work in other works.}}

\begin{document}

\title{REGATTA: A Cross-Domain Measurement Framework and
Fault-Injection Harness for Enterprise AI Agents}

\author{\IEEEauthorblockN{Luong Tuan Thanh\,\orcidlink{0009-0000-1199-837X}}
\IEEEauthorblockA{Golden Gate University, San Francisco, CA, USA\\
tluongtuan@my.ggu.edu}
\and
\IEEEauthorblockN{Abhijit Sanyal\,\orcidlink{0009-0005-7520-5881}}
\IEEEauthorblockA{Data, Digital \& IT, Novartis Healthcare Pvt.\ Ltd., Hyderabad, India\\
abhijit.sap@gmail.com}}

\maketitle

\begin{abstract}
Enterprise AI-agent evaluation needs evidence about whether test suites
expose specified faults and whether their reference rules are trustworthy.
We introduce REGATTA (Regulatory Evaluation of Agents via Two-stage
Testing and Assurance), a measurement framework with three instruments.
A 125-item source-cited checklist spans five industry-by-jurisdiction
cells across the United States and Vietnam. A 25-fault corpus covers
five categories, and a harness scores suite-level fault detection at
the design and execution stages. We examine four measurement
diagnostics on 180 generated suites (5{,}400 scenarios) from three
LLM generators. A source audit finds 24 items misstating the rule in
force, 46 additional citation defects, 31 items with only non-binding,
unidentifiable, or unverifiable sources, and 24 verified clean.
Only initially substance-flagged items received author review; all
24 clean rows received AI review only. Results therefore describe
the instruments as written, not legal compliance. Item coverage
varies, execution-stage fault profiles are stable across generators,
and the historical stages agree on $66$--$69\%$ of fault--suite pairs.
A current-model US-insurance retest finds stage disagreement above
repeated-execution variation, but stage agreement $.536$ falls below
pooled chance $.549$, failing V3's lower inequality.
Blind adjudication by one expert supports the design-stage judging
protocol ($\kappa=.72$, 95\% interval $[.52,.88]$); 40 current-model
execution probes give $\kappa=.75$. The instruments, audit, and
reusable code are released.
\end{abstract}


\section{Introduction}
\label{sec:intro}

Banks, insurers, and hospitals need evidence that agent tests expose
failures against specific obligations. Agent capability suites such as
AgentBench~\cite{liu2023agentbench} and $\tau$-bench~\cite{yao2024taubench}
measure task success and policy following, but those scores do not
establish a suite's adequacy against a jurisdiction's stated rules.
The NIST AI Risk Management Framework calls for
measurable assurance evidence~\cite{nist2023rmf}. Vietnam's AI Law,
in force only since March 1, 2026, requires conformity assessment of
high-risk AI systems before deployment and leaves the list of such
systems to the Prime Minister~\cite{vnailaw2025}.

A scenario can mention a rule without exposing a
running agent's violation. This paper studies that gap, packaged
as \textbf{REGATTA} (Regulatory Evaluation of Agents via Two-stage
Testing and Assurance), with an audit of their own legal sources:

\begin{itemize}
  \item \textbf{C1 --- Checklists and source audit:} 125
    binary-scored requirement items in five industry-by-jurisdiction
    cells, two under Vietnamese law, a low-resource regime. Each item
    cites its source, and every citation is audited against primary
    records.
  \item \textbf{C2 --- Fault corpus:} 25 seeded agent faults (five per
    cell, five categories), each with a stated correct behavior and
    detection mechanism, after mutation
    analysis~\cite{jia2011mutation}; audited like C1.
  \item \textbf{C3 --- Two-stage harness:} design- and execution-stage
    fault detection plus coverage, specificity, and adversarial
    LLM judges~\cite{zheng2023judging}, scored per suite.
  \item \textbf{C4 --- Measurement model:} four measurement diagnostics
    (Section~\ref{sec:formal}), examined on 180 suites under three
    generators, with second-vendor sensitivity checks on selected
    measures (Section~\ref{sec:validation}).
\end{itemize}

REGATTA combines suite-level coverage, predicted fault exposure,
execution observations, and an item-level legal-source audit. Its
contribution is this measurement protocol and the comparison of its
stages across five cells, building on mutation-based test adequacy
and stochastic agent testing. The audit requires correcting C1 and
C2 before legal use. We examine operational properties, not universal
validity standards or which generation strategy is superior. Blind expert review was performed for the original design/coverage
protocols and 40 current-model execution probes by the same expert;
a second-vendor judge provides a sensitivity check (\S\ref{sec:kappa}).

\IEEEpubidadjcol{}
\section{Related Work}
\label{sec:related}

\subsection{Agent Benchmarks}

AgentBench measures interactive task completion~\cite{liu2023agentbench};
$\tau$-bench also tests domain-policy following and repeated-run
reliability~\cite{yao2024taubench}. AgentHarm measures harmful-request
susceptibility~\cite{andriushchenko2024agentharm}, and LegalBench
predominantly tests American legal reasoning~\cite{guha2023legalbench}.
AIR-Bench derives safety categories from regulations and
policies~\cite{zeng2025airbench}. These establish that policy-grounded
evaluation itself is not new.

Mutation testing scores suites using seeded
defects~\cite{jia2011mutation}. ACH generates tests against
concern-specific program mutants~\cite{harman2025ach}.
ToolEmu uses an LLM-emulated sandbox and an automatic safety
evaluator to identify agent failures, with human checks of the
emulator and evaluator~\cite{ruan2024toolemu}.
REGATTA's narrower focus is paired predicted and observed exposure of
prompt-configured faults, alongside an audit of the source-cited
instruments. Unlike controlled mutant comparisons, it does not
isolate mutation effects from ordinary failures or fault interactions.

\subsection{Governance Frameworks and Enterprise Evaluation}
\label{sec:governance}

The NIST AI RMF organizes AI risk management~\cite{nist2023rmf},
while GoldMine OS proposes an agent architecture for alternative-asset
trading~\cite{borjigin2025goldmine}. ShieldAgent checks action
trajectories against explicit safety policies using verifiable
reasoning~\cite{chen2025shieldagent}. AgentSpec supplies executable
runtime constraints~\cite{wang2026agentspec}, complementary to
semantic judging. REGATTA examines scenario-suite adequacy; these
systems have different evaluation units, so their scores are not
numerical baselines.
Ontology-structured context follows
\cite{luong2026neurosymbolic} for strategies S3 and S4.

\section{The REGATTA Framework}
\label{sec:framework}
\label{sec:formal}

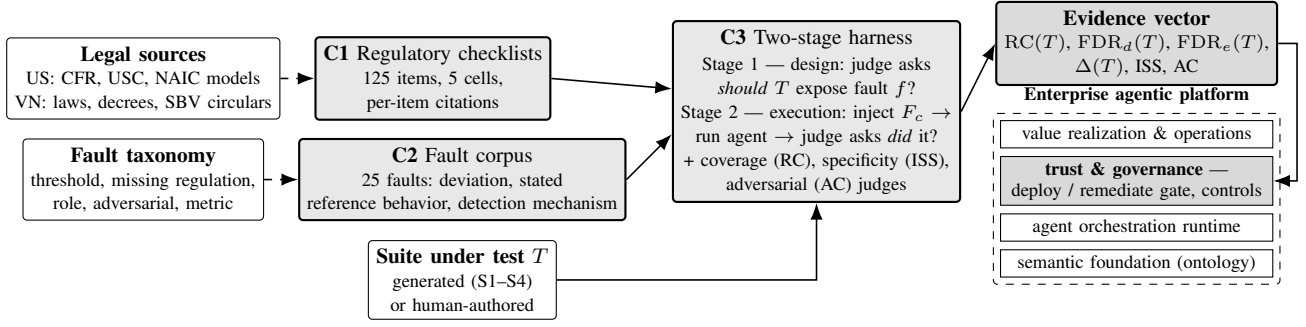
\begin{figure*}[t]
\centering
\begin{tikzpicture}[
  font=\footnotesize,
  node distance=2.5mm and 4.5mm,
  box/.style={draw, rounded corners=1.5pt, align=center,
              inner sep=3pt, minimum height=9mm},
  src/.style={box, fill=white},
  inst/.style={box, fill=gray!18, thick},
  harn/.style={box, fill=gray!18, thick},
  outp/.style={box, fill=gray!28, thick},
  pl/.style={draw, align=center, font=\scriptsize, minimum width=36mm,
             inner sep=2pt, fill=white},
  arr/.style={-{Latex[length=2.2mm]}, semithick},
  cur/.style={-{Latex[length=2.2mm]}, semithick, dashed}
]
\node[src] (statutes) {\textbf{Legal sources}\\
  {\scriptsize US: CFR, USC, NAIC models}\\
  {\scriptsize VN: laws, decrees, SBV circulars}};
\node[src, below=of statutes] (taxonomy) {\textbf{Fault taxonomy}\\
  {\scriptsize threshold, missing regulation,}\\
  {\scriptsize role, adversarial, metric}};
\node[inst, right=of statutes] (c1) {\textbf{C1} Regulatory checklists\\
  {\scriptsize 125 items, 5 cells,}\\
  {\scriptsize per-item citations}};
\node[inst, right=of taxonomy] (c2) {\textbf{C2} Fault corpus\\
  {\scriptsize 25 faults: deviation, stated}\\
  {\scriptsize reference behavior, detection mechanism}};
\node[src, below=of c2] (suite) {\textbf{Suite under test} $T$\\
  {\scriptsize generated (S1--S4)}\\
  {\scriptsize or human-authored}};
\node[harn, right=6mm of c2, yshift=9mm]
  (c3) {\textbf{C3} Two-stage harness\\[1pt]
  {\scriptsize Stage 1 --- design: judge asks}\\
  {\scriptsize \emph{should} $T$ expose fault $f$?}\\
  {\scriptsize Stage 2 --- execution: inject $F_c$ $\rightarrow$}\\
  {\scriptsize run agent $\rightarrow$ judge asks \emph{did} it?}\\
  {\scriptsize + coverage (RC), specificity (ISS),}\\
  {\scriptsize adversarial (AC) judges}};
\node[outp, right=of c3, yshift=9mm] (ev) {\textbf{Evidence vector}\\
  {\scriptsize $\mathrm{RC}(T)$, $\mathrm{FDR}_{d}(T)$,
    $\mathrm{FDR}_{e}(T)$,}\\
  {\scriptsize $\Delta(T)$, ISS, AC}};
\node[pl, below=4.5mm of ev] (l4) {value realization \& operations};
\node[pl, fill=gray!30, below=1mm of l4] (l3)
  {\textbf{trust \& governance} ---\\deploy / remediate gate, controls};
\node[pl, below=1mm of l3] (l2) {agent orchestration runtime};
\node[pl, below=1mm of l2] (l1) {semantic foundation (ontology)};
\node[draw, dashed, rounded corners=1.5pt, fit=(l4)(l1),
  inner sep=2.5pt,
  label={[font=\scriptsize\bfseries]above:%
    Enterprise agentic platform}] (platform) {};
\draw[cur] (statutes) -- (c1);
\draw[cur] (taxonomy) -- (c2);
\draw[arr] (c1.east) -- ([yshift=3mm]c3.west);
\draw[arr] (c2.east) -- ([yshift=-3mm]c3.west);
\draw[arr] (suite.east) -| (c3.south);
\draw[arr] (c3.east) -- (ev.west);
\draw[arr] (ev.east) -- ++(2.5mm,0) |- (l3.east);
\end{tikzpicture}
\caption{REGATTA architecture. Legal sources and a fault taxonomy
are curated into two static instruments (C1, C2; dashed arrows:
one-time curation, checked by the source audit). The two-stage harness (C3)
scores any test suite $T$, generated (S1--S4) or human-authored, and
emits an evidence vector for governance review in the deploying
platform~\cite{nist2023rmf}. RC:
requirement coverage; FDR$_d$, FDR$_e$: design- and execution-stage
fault-detection rates; ISS: industry specificity (1--5); AC:
adversarial-category coverage.}
\label{fig:architecture}
\end{figure*}

Fig.~\ref{fig:architecture} separates source-cited requirement
statements (C1), seeded fault descriptions (C2), and scoring
procedures (C3), so each can extend independently. The evidence vector summarizes suite behavior for a subsequent
assurance process; it is not a deployment decision rule.

A \emph{cell} $c = (d, j)$ pairs an industry domain with a
jurisdiction and supplies a requirement set
$R_c$ ($|R_c| = 25$) and a fault set $F_c$ ($|F_c| = 5$); a test
suite $T = \{s_1, \dots, s_n\}$ is a set of scenarios ($n = 30$
throughout). Three binary judge functions --- fixed rubrics at
temperature $0$ --- generate the detection measurements:
$J_{\mathrm{RC}}(s, r)$ asks whether scenario $s$ tests requirement
$r$; $J_{D}(s, f)$ asks whether $s$, as specified, should expose
fault $f$; and $J_{E}(\pi_{F_c}(s), f)$ asks whether the response
$\pi_{F_c}(s)$ of the agent with the cell's faults injected actually
surfaces $f$. The suite-level measurements are maxima over scenarios,
normalized to rates:
\begin{align}
\mathrm{RC}(T) &= \tfrac{1}{|R_c|} \textstyle\sum_{r \in R_c}
  \max_{s \in T} J_{\mathrm{RC}}(s, r),\\
\mathrm{FDR}_{d}(T) &= \tfrac{1}{|F_c|} \textstyle\sum_{f \in F_c}
  D(T, f),\\
\mathrm{FDR}_{e}(T) &= \tfrac{1}{|F_c|} \textstyle\sum_{f \in F_c}
  E(T, f),
\end{align}
with $D(T,f) = \max_{s \in T} J_{D}(s, f)$ and
$E(T,f) = \max_{s \in T} J_{E}(\pi_{F_c}(s), f)$.
The faulted agent $\pi_{F_c}$
samples at temperature $0.3$, so each probe inside $E(T,f)$ rests on
one agent draw (Section~\ref{sec:v3} reports a retest).
These are mutation-inspired fault-exposure measures: design predicts
exposure; execution records judge-assessed exposure under joint fault
injection. Without an unmodified-agent control, neither isolates a
mutation effect. Their difference is the suite's \emph{stage
divergence}, $\Delta(T) = \mathrm{FDR}_{d}(T) - \mathrm{FDR}_{e}(T)$,
which the harness reports as a first-class quantity rather than a
nuisance. Two further protocols score industry specificity
(ISS, $1$--$5$) and adversarial-category coverage (AC); both enter
the evidence vector and V4 audits their discrimination. Over a
population of suites
$\mathcal{T}$, each requirement
has an empirical \emph{coverage rate}
$\delta(r) = \mathbb{E}_{T \in \mathcal{T}}[\max_{s \in T}
J_{\mathrm{RC}}(s, r)]$ --- low-$\delta$ items form the checklist's
hard tail --- and the two stages have an \emph{agreement
rate} $\alpha = \Pr[D(T,f) = E(T,f)]$ over fault--suite pairs, with
chance baseline $\alpha_0 = pq + (1-p)(1-q)$ for marginal detection
rates $p = \Pr[D{=}1]$, $q = \Pr[E{=}1]$. For a measure $m$, the
between-strategy range is $\rho_m = \max_{\sigma} \bar{m}(\sigma) -
\min_{\sigma} \bar{m}(\sigma)$ over generation strategies $\sigma$.

The four measurement diagnostics (V1--V4) are evaluation questions.
Each links a procedure to its result; V1 and V4 remain descriptive
because their practical thresholds were not fixed in advance.

\begin{itemize}
  \item \textbf{V1 --- Item-coverage spread:} do coverage rates vary
    across requirements? Procedure: $\delta(r)$ over the 36 suites
    in each item's cell (\S\ref{sec:v1}). Diagnostic: dispersion,
    tails, and cell saturation, also checked under GPT-4o. This
    describes items; suite separation is examined in V4.
  \item \textbf{V2 --- Profile stability:} is the execution profile
    stable across the tested generators? Procedure: category-level
    $\mathrm{FDR}_e$ per generator (\S\ref{sec:v2}). Diagnostic: the
    top-three order and bottom-pair membership are invariant; inversions would limit
    transfer across generators.
    Design-stage profiles may vary. The original judge yields this
    ordering across the tested generators.
  \item \textbf{V3 --- Stage non-redundancy:} do the stages carry
    distinct information? Procedure: $\alpha$ over 300 fault--suite
    pairs per generator (\S\ref{sec:v3}). Threshold: $\alpha_0 <
    \alpha$, as a positive-association check. In a cell retested within one window, the stage
    agreement measured there must also stay below $\alpha_{EE} =
    \Pr[E_1(T,f) = E_2(T,f)]$, to require excess stage
    disagreement over observed retest disagreement; elsewhere $\alpha < 1$. If the lower 95\% bound of $\alpha_{EE}$ reaches
    $.98$, that upper bound is uninformative. Historical agreement
    clears pooled chance, but the 26B per-fault margin spans zero.
    In the retest, $\alpha < \alpha_{EE}$, but $.536 < .549$ fails
    the pooled lower inequality; the per-fault margin also spans zero.
  \item \textbf{V4 --- Discrimination:} does each core measure (RC,
    both FDRs, ISS) separate strategies? Procedure: $\rho_m$ per
    generator (\S\ref{sec:v4}). Diagnostic: observed separation under every generator; no practical
    floor or interval was fixed, so this is not a significance test.
    Core $\rho_m \ge .08$ ($\ge .07$ under GPT-4o); AC nearly
    saturates under two generators.
\end{itemize}

\section{The Instruments}
\label{sec:design}

The judging rules are curated separately from the ontology context
given to the generator; semantic overlap and shared model assistance
can still introduce correlated errors.
The cells in Table~\ref{tab:cells} cover the US and Vietnam.

\begin{table}[t]
\centering
\caption{Checklist cells and principal sources as cited at run time
(25 items each).}
\label{tab:cells}
\begin{tabular}{@{}lll@{}}
\toprule
\textbf{Cell} & \textbf{Regime} & \textbf{Sources as cited} \\
\midrule
Fintech & US & 31 CFR \S\S1010--1020; OCC 2013-29; \\
        &    & Dodd-Frank \S1071 \\
Insurance & US & NAIC Models \#785/\#668/\#900; \\
          &    & RBC instructions; state DOI filings \\
Healthcare & US & 45 CFR \S164; 42 USC \S1395dd; \\
           &    & 42 CFR \S411 \\
Banking & VN & SBV Circulars 11/2021, 22/2019, \\
        &    & 39/2016; Decree 116/2013 \\
Insurance & VN & Insurance Business Law 2022; \\
          &    & Circular 67/2023; Decree 46/2023; \\
          &    & AML Law 2022~\cite{vnamllaw2022} \\
\bottomrule
\end{tabular}\\[2pt]
\parbox{0.97\columnwidth}{\footnotesize Several sources were superseded
or amended before the runs; the source audit checks every citation (C1).}
\end{table}

\begin{itemize}
  \item \textbf{C1 --- checklists:} 25 items per cell, each one
    testable obligation (\emph{file a currency-transaction report for
    cash above \$10{,}000}) with identifier, source, requirement
    text, and category. Claude, from the judge's vendor, drafted some
    checklist items, fault records, and ontology passages, which the
    authors revised. After the runs, AI agents audited every citation
    against primary records for the run window (March 29 to April 11,
    2026), with an adversarial cross-check. The first author reviewed
    every row flagged as a substance error, using an AI search tool
    to locate sources and deciding each row after reading them. After
    that review, a complete disjoint partition of 125 items is:
    24 substance errors; 46 additional superseded or incorrect
    citations; 17 non-binding or unidentifiable-source-only items;
    3 unverifiable-only items; 11 with both source flags; and
    24 verified clean. Substance errors take priority over citation
    defects, then the residual source flags, then clean status.
    The 30 initially substance-flagged items received author review.
    The other 95, including all 24 clean rows, received AI review
    only, with no human review. Across these classes, 45 rest on
    non-binding or unidentifiable sources. Item texts enter the judge prompts, so every result
    measures the instruments as written. The item-level audit
    marks C1 and C2 as requiring correction
    before use as legal ground truth.
  \item \textbf{C2 --- faults:} one per category and cell: threshold
    errors (a \$15{,}000 reporting threshold where the rule says
    \$10{,}000), missing regulations, role-boundary violations,
    adversarial vulnerabilities, and metric errors. Each record holds
    the injected deviation, ground-truth behavior, severity, and
    detection mechanism. A cell's five faults are injected together
    into the system prompt the agent runs under, never into the
    scenarios. Seven of the 25 misstate their rule. In two Vietnamese
    faults (a widened loan-classification window, a solvency margin at 80\% of the
    floor) the injected deviation remains an error, but the
    ground-truth text is partly wrong.
  \item \textbf{C3 --- harness:} it now includes a deterministic holdout
    seed and per-suite manifest, but the 180 suites predate them.
    A July 2026 code audit
    found the earlier routine withheld nothing in three cells and one
    or two items per suite in two, with no manifest. S4 is therefore
    read as structured-ontology prompting with no effective holdout.
\end{itemize}

\section{Validation and Release}
\label{sec:validation}

\subsection{Setup}

We generated 180 suites of 30 scenarios (5{,}400 scenarios): four
strategies $\times$ five cells $\times$ three replications, each an
independently generated suite (temperature $0.3$). Three generators
produced 60 suites each: Claude Sonnet~4 (\emph{frontier}; also the
judge and the faulted agent), Qwen~2.5 72B Instruct (\emph{72B}), and
Gemma~4 26B-A4B IT (\emph{26B}). The strategies are minimal context
(S1), persona-matrix prompting as commercial simulators use (S2),
retrieval over unstructured ontology text (S3), and structured
ontology grounding (S4)~\cite{luong2026neurosymbolic}; no analysis
tests which is superior. One faulted agent is exercised through
single-turn scenarios, so generator invariance is tested (V2) but
agent invariance is not. Execution-stage profiles and stage agreement
may shift with another agent, which the retest of \S\ref{sec:v3}
begins to probe on one cell.

\subsection{Checklist Difficulty Spread (V1)}
\label{sec:v1}

Pooled equal-weight across generators, mean item coverage $\delta$ is
$.34$ (median $.28$, SD $.31$): 5 items are covered by over 90\% of
suites, 60 by under a quarter, 22 by none. The never-covered tail sits
in consumer-protection disclosures (APR, adverse action), prudential
ratios (CET1, LCR, DTI), and administrative duties, outside the
anti-money-laundering and privacy core all strategies reach. Per cell,
mean coverage runs from $.28$ (Insurance US) to $.42$ (Banking VN), so
no cell saturates in this sample (V1). Fintech is nearly bimodal, with 12 of
25 items never covered under either judge. Its pooled detection falls
from design to execution ($.56 \rightarrow .34$), while both
Vietnamese cells rise ($.48 \rightarrow .58$, $.41 \rightarrow .56$),
so results should be read per cell.

\subsection{Fault Detectability Profiles (V2)}
\label{sec:v2}

\begin{table}[t]
\centering
\caption{Fault-category detection rates: design stage $\rightarrow$
execution stage, per generator model (five faults per category).}
\label{tab:faultcat}
\setlength{\tabcolsep}{4.5pt}
\begin{tabular}{@{}lccc@{}}
\toprule
\textbf{Category} & \textbf{Frontier} & \textbf{72B} & \textbf{26B} \\
\midrule
Threshold error    & $.75 \rightarrow .63$ & $.32 \rightarrow .72$ & $.62 \rightarrow .57$ \\
Missing regulation & $.87 \rightarrow .95$ & $.78 \rightarrow .98$ & $.82 \rightarrow .95$ \\
Role boundary      & $.45 \rightarrow .32$ & $.17 \rightarrow .35$ & $.32 \rightarrow .18$ \\
Adversarial vuln.  & $.67 \rightarrow .20$ & $.30 \rightarrow .17$ & $.70 \rightarrow .28$ \\
Metric error       & $.20 \rightarrow .38$ & $.22 \rightarrow .40$ & $.18 \rightarrow .43$ \\
\bottomrule
\end{tabular}
\end{table}

Missing-regulation faults approach ceiling at execution
($.95$--$.98$; Table~\ref{tab:faultcat}): when a mandated capability
is absent, almost any on-topic scenario exposes it. Adversarial faults
collapse from design to execution under all three generators
($.67\!\rightarrow\!.20$, $.30\!\rightarrow\!.17$,
$.70\!\rightarrow\!.28$). Suites contain scenarios that should probe
the weakness, yet those rarely trigger it when run. Metric-error
faults invert: hard to cover by design ($.18$--$.22$; at most $.28$
under GPT-4o), they surface at execution ($.38$--$.43$) because a
faulty calculation leaks into unrelated responses. At execution, the
top-three order (missing regulation $>$ threshold $>$ metric) is
identical under all three generators, and role-boundary and
adversarial faults form the floor pair (every rate $\le .35$),
satisfying V2's ordering diagnostic in this sample. Design-stage orders vary (the 72B swaps metric and role under
both judges).

\subsection{Stage Non-Redundancy (V3)}
\label{sec:v3}

Design and execution verdicts disagree on $32.3\%$ (frontier),
$30.7\%$ (72B), and $33.7\%$ (26B) of fault--suite pairs. Against
chance baselines from each generator's marginal detection rates
($\alpha_0 = .499$, $.493$, $.499$), agreement is $\alpha = .677$,
$.693$, and $.663$. Its 95\% intervals resample suites within cell (10,000 draws),
keeping faults together:
$[.63, .72]$, $[.66, .73]$, and $[.62, .71]$, and the margins over
$\alpha_0$ span $[.13, .22]$, $[.17, .23]$, and $[.12, .21]$.
These clear the pooled V3 baseline; the per-fault result is weaker. Per-fault baselines ($.648/.653/.646$) are cleared by small
margins ($+.029/+.040/+.017$). Their intervals exclude zero for the
frontier and 72B but not the 26B ($[-.008, .038]$), so most stage
agreement tracks fault difficulty.

Execution noise could explain part of the historical disagreement:
each verdict used one agent draw at temperature $0.3$, and no responses
were stored. A current-model retest uses Claude Sonnet~4.6 as faulted
agent and judge with the frozen harness. A rule fixed before research
calls selected US insurance: 36 existing suites, 180 fault--suite
pairs, a same-window design re-scan, and two execution draws with all
outputs retained. This is not a re-measurement of the original model.
The cell uses model-act, internal-policy, and technical reference
rules; it does not establish compliance with a state's enacted law.

Table~\ref{tab:validation} shows stage disagreement exceeding execution
variation, satisfying V3's upper inequality. Stage agreement $.536$
falls below pooled chance $.549$, so the lower inequality fails.
Against the per-fault baseline $.522$, the margin $.014$
$[-.019,.046]$ also does not establish positive association.
The interpretation of registered reading A was narrowed after analysis
to excess stage disagreement, rather than full V3 satisfaction.
Post-analysis Addendum~3 (September 28, 2026) records this narrowing
without changing the registered estimands or reading rule~\cite{regatta2026protocol}.
The re-scan matched 143/180 historical verdicts ($.794$), mixing model
change with judging variance; there were no strict-parse failures.

\begin{table}[t]
\centering
\caption{Current-model validation in US insurance. Rows 1--3 use 180
matched fault--suite pairs; row 4 uses 7,816 probes; row 5 uses 40
balanced probes. 95\% intervals: 10,000 suite-bootstrap draws within
generator, keeping faults and both draws together (rows 1--4);
verdict-stratified bootstrap (row 5).}
\label{tab:validation}
\setlength{\tabcolsep}{3pt}
\begin{tabular}{@{}lcc@{}}
\toprule
\textbf{Measure} & \textbf{Estimate} & \textbf{95\% interval} \\
\midrule
Repeat-execution agreement & $.917$ & $[.878,.950]$ \\
Stage agreement, draw average & $.536$ & $[.475,.597]$ \\
Excess stage disagreement & $.381$ & $[.317,.447]$ \\
Sonnet--GPT-4o $\kappa$ & $.621$ & $[.556,.695]$ \\
Sonnet--expert $\kappa$ & $.750$ & $[.550,.900]$ \\
\bottomrule
\end{tabular}
\end{table}

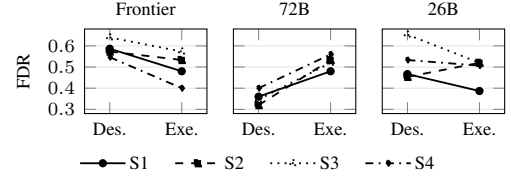
\begin{figure}[t]
\centering
\begin{tikzpicture}
\begin{groupplot}[
  group style={group size=3 by 1, horizontal sep=0.35cm,
               ylabels at=edge left, yticklabels at=edge left},
  width=3.2cm, height=2.7cm,
  xmin=-0.35, xmax=1.35, ymin=0.28, ymax=0.68,
  xtick={0,1}, xticklabels={Des., Exe.}, tick label style={font=\scriptsize},
  ytick={0.3,0.4,0.5,0.6}, ymajorgrids, grid style={gray!22, line width=0.2pt},
  scaled ticks=false, every axis plot/.append style={line width=0.7pt, mark size=1.2pt},
  title style={font=\scriptsize, yshift=-1ex},
]
\nextgroupplot[title={Frontier}, ylabel={\scriptsize FDR}]
\addplot[solid,      mark=*]           coordinates {(0,0.5867)(1,0.4800)};
\label{pgf:s1}
\addplot[dashed,     mark=square*]     coordinates {(0,0.5733)(1,0.5333)};
\label{pgf:s2}
\addplot[dotted,     mark=triangle, mark size=1.5pt]
                                       coordinates {(0,0.6400)(1,0.5733)};
\label{pgf:s3}
\addplot[dashdotted, mark=diamond*]    coordinates {(0,0.5467)(1,0.4000)};
\label{pgf:s4}
\nextgroupplot[title={72B}]
\addplot[solid,      mark=*]           coordinates {(0,0.3600)(1,0.4800)};
\addplot[dashed,     mark=square*]     coordinates {(0,0.3200)(1,0.5333)};
\addplot[dotted,     mark=triangle, mark size=1.5pt]
                                       coordinates {(0,0.3467)(1,0.5200)};
\addplot[dashdotted, mark=diamond*]    coordinates {(0,0.4000)(1,0.5600)};
\nextgroupplot[title={26B}]
\addplot[solid,      mark=*]           coordinates {(0,0.4667)(1,0.3867)};
\addplot[dashed,     mark=square*]     coordinates {(0,0.4533)(1,0.5200)};
\addplot[dotted,     mark=triangle, mark size=1.5pt]
                                       coordinates {(0,0.6533)(1,0.5200)};
\addplot[dashdotted, mark=diamond*]    coordinates {(0,0.5333)(1,0.5067)};
\end{groupplot}
\end{tikzpicture}
\\[-0.6ex]
{\scriptsize
\ref{pgf:s1}\,S1 \quad \ref{pgf:s2}\,S2 \quad
\ref{pgf:s3}\,S3 \quad \ref{pgf:s4}\,S4}
\caption{Design-stage vs.\ execution-stage fault-detection rate (FDR)
by generation strategy (S1--S4, \S\ref{sec:validation}), per generator
model, as scored by the original judge.}
\label{fig:signflip}
\end{figure}

The direction of disagreement depends on the generator
(Fig.~\ref{fig:signflip}). Under the original judge, every strategy's
execution score falls below its design score for the frontier
generator and exceeds it for the 72B, while the 26B mixes signs. The
frontier's uniform sign does not survive GPT-4o design verdicts.
Pooled $\Delta$ is $+.09$ $[.05, .13]$, $-.17$ $[-.21, -.12]$, and
$+.04$ $[-.01, .09]$, with each pooled stage rate known within
$\pm .04$. For a deployment gate, $\mathrm{FDR}_e$ is the
operative evidence, the seeded faults a suite exposed in a running
agent, and $\Delta(T)$ is how far a design-only review would misstate
it. Both concern seeded faults in one agent configuration, not
production violation rates.

\subsection{Strategy Discrimination (V4)}
\label{sec:v4}

\begin{table}[t]
\centering
\caption{Between-strategy separation $\rho_m$ (range of suite-level
means over the four strategies), per measure and generator model.}
\label{tab:discrim}
\setlength{\tabcolsep}{4.5pt}
\begin{tabular}{@{}lccc@{}}
\toprule
\textbf{Measure} & \textbf{Frontier} & \textbf{72B} & \textbf{26B} \\
\midrule
Requirement coverage {\footnotesize(0--1)}   & $.15$ & $.14$ & $.16$ \\
Design-stage FDR {\footnotesize(0--1)}       & $.09$ & $.08$ & $.20$ \\
Execution-stage FDR {\footnotesize(0--1)}    & $.17$ & $.08$ & $.13$ \\
Industry specificity {\footnotesize(1--5)}   & $.24$ & $.53$ & $.38$ \\
Adversarial coverage {\footnotesize(0--1)}   & $.03$ & $.48$ & $.03$ \\
\bottomrule
\end{tabular}
\end{table}

Every core measure shows observed separation under every generator
(Table~\ref{tab:discrim}): coverage by 14--16 percentage points,
specificity by $.24$--$.53$ Likert points, and the detection rates by
$.08$--$.20$. These ranges are descriptive: without uncertainty
intervals or a prespecified floor, V4 is not inferential validation.
Adversarial-category coverage is weak: under the frontier and 26B
generators all four strategies near-saturate the six-category
taxonomy, so it needs finer categories.

\subsection{Judge Agreement}
\label{sec:kappa}

A financial expert with at least 10 years of experience
blindly adjudicated 75 pairs from the persisted
verdicts (all five cells, both open-weights generators' suites),
answering the judges' own questions without AI tools. The first
author transcribed the answers. Design stage: 50 fault--scenario
pairs, 25 judge-detected via the persisted detecting scenario and 25
judge-undetected, every pair verdict-recoverable through the scan's
early-exit semantics. Agreement is $.86$ (Cohen's $\kappa = .72$;
95\% CI $[.52, .88]$ from a bootstrap within the two verdict strata).
All seven disagreements run one way, judge-credited and
expert-rejected, typically where a scenario touches the fault's topic
without forcing faulty and correct behavior to differ observably. No
judge-undetected pair was overturned. This sample suggests design-judge leniency, which can inflate
design-to-execution drops and understate rises; it does not establish
a population-wide upper bound on detection. The expert
also audited 25 judge-uncovered requirement pairs --- the verdict
class the never-covered-tail analysis rests on --- and confirmed all
25 (exact 95\% lower bound $.86$). This supports the sampled negatives, not the entire hard tail. These checks provide no independent inter-expert agreement estimate. The same expert, blinded and unaided, assessed 40 retest probes (20 judge-positive, 20 judge-negative): agreement $.875$ (Table~\ref{tab:validation}). All five disagreements were judge-positive/expert-negative. A separate two-expert discussion panel returned identical verdicts; individual panel labels were not supplied. These balanced-sample rates are not population accuracies. An AI-assisted review was excluded from the human check.

OpenAI's GPT-4o (\texttt{gpt-4o-2024-11-20}) re-scored coverage, design-stage detection, AC, and ISS on all 180
suites, given byte-identical prompts at temperature $0$. Against the
original judge, its raw agreement on coverage items is $.862$, $.890$,
and $.875$ ($\kappa = .705$, $.647$, $.723$) for the frontier, 72B,
and 26B suites. On design-stage verdicts it is $.897$, $.843$, and
$.873$ ($\kappa = .789$, $.666$, $.746$). Chance-corrected agreement is
lower on ISS (quadratic-weighted $\kappa = .42$--$.62$) and on AC, which sits near ceiling
under two generators ($\kappa = .23$--$.64$ at raw agreement
\mbox{$.84$--$.96$}). Recomputed with its verdicts
(for V3, its design verdicts against the original execution verdicts),
V1 and V4 retain their descriptive patterns. Mixed-judge V3 clears
pooled and per-fault baselines, but confounds judge and stage
differences; the original 26B per-fault interval spans zero. GPT-4o is stricter on coverage
(pooled mean $.25$ against $.34$) but ranks item difficulty alike
(Spearman $.90$). V2 rests on execution verdicts, which only the
original judge scored in the published runs. On 7,816 stored retest probes, GPT-4o agrees with Sonnet~4.6 on $.986$; its $\kappa$ interval (Table~\ref{tab:validation}) gives an intermediate result under the pre-registered thresholds. GPT-4o uses registered code-fence removal; strict parsing leaves that classification unchanged. Both judges label 7,617 probes negative, inflating raw agreement. On the 40-probe expert sample, GPT-4o agrees on 31 ($\kappa=.50$). The check follows Sonnet's early-stopping scans and covers one fault per category; it assesses judge agreement, not V2's ordering.

\subsection{Evaluation Cost}
\label{sec:cost}

One 30-scenario suite costs at most
$1{,}110$ judge calls ($750$ RC, $60$ ISS and AC, $150$ design-stage,
$150$ execution) and $150$ agent calls. Scans stop at their first
success, so realized cost falls as suite quality rises. A fresh agent
response per fault and scenario keeps the five verdicts from sharing
a draw.

\subsection{Availability and Extension}
\label{sec:availability}

The research archive (v1.1)~\cite{regatta2026archive} includes all
5{,}400 recovered scenarios, the instruments and source audit,
adapted code, selected original verdicts, 4{,}313 design judgments,
7{,}816 masked execution probes, and 40 deidentified human labels.
Research code uses the MIT license; instruments and derived data use
CC BY~4.0. Offline checks reproduce 2{,}835 original descriptive
reference values and 83 selected current-model values; the inventory
states exclusions. Historical agent responses and holdout memberships
were not retained. The retest of \S\ref{sec:v3} retained its responses
and verdicts, but masking changes prompt bytes. Exact replay is not
claimed.
A new cell follows a six-step protocol. Enumerate the
governing sources; curate 25 observably-testable obligations with
citations; author five faults with ground truth and detection
mechanism; audit every citation against the instruments in force;
run the harness unchanged; and re-examine diagnostics V1--V4.

\section{Conclusion}
\label{sec:conclusion}

REGATTA supplies a measurement framework over source-cited
checklists and specified faults. Four diagnostics examine coverage
spread, profile stability, stage agreement, and strategy separation.
Across 180 suites, V1 and V4 show descriptive variation and V2's
execution-profile ordering is stable across the tested generators.
Historical stage agreement exceeds pooled chance, but the original
26B per-fault margin spans zero ($[-.008,.038]$).
In the Sonnet~4.6 retest, $.536 < .549$ fails V3's pooled lower
inequality; the per-fault margin $.014$ $[-.019,.046]$ also spans zero.
Excess stage disagreement remains positive, with reading A narrowed
post hoc to that result~\cite{regatta2026protocol}.
Blind expert and second-vendor checks qualify judging uncertainty.
The evidence concerns specified faults under this protocol, not legal
compliance. Source correction and independent review are required
before practical assurance use.

Five limitations bound the scope of these results. First, the
evaluation relies on LLM-based judging, checked by a single expert.
That expert's review gave $\kappa = .72$ (95\% CI $[.52, .88]$) on
design-stage judgments and upheld all 25 audited coverage negatives;
40 current-model execution probes give $\kappa=.75$, with no independent inter-expert estimate.
Because the published runs did not persist agent responses, their
execution verdicts cannot be re-adjudicated from the release. Second,
one faulted agent was exercised, through single-turn scenarios with
all of a cell's faults injected at once, so the execution-stage
results (V2, V3) may not transfer to other agents. The instrument set also
remains relatively small, with five faults and 25 requirements per
cell. This resolution facilitates sampling a jurisdiction's
obligations rather
than providing exhaustive coverage; deployment use requires corrected rules, independently validated
oracles, and externally justified decision thresholds. Third, the ground truth is
author-curated and failed its own audit: 24 of 125 items misstate
the rule in force, 45 rest on non-binding or unidentifiable sources,
and no lawyer reviewed the audit. The reported measurements describe
the instruments as written, not statements of law. Fourth, the
potential for judge--generator affinity exists, as one generator
shares the same vendor as the judge, which may lead to inflated
scores for that model in Tables~\ref{tab:faultcat}
and~\ref{tab:discrim} and Fig.~\ref{fig:signflip}. The same vendor's
model also helped draft the instruments, a second route to affinity.
Finally, the adversarial taxonomy appears to saturate under two of
the three generators, effectively separating strategies only under
the third.

Future work will include correcting the instruments with domain
experts before legal use, obtaining independent multi-expert execution
labels, and extending repeated execution and second-judge checks
beyond one cell. After those steps, denser fault sets, additional
jurisdictions, and finer adversarial categories will be evaluated
with prespecified thresholds.

\section*{Generative AI Usage Acknowledgment}

The generators, the faulted agent, both judges (Claude Sonnet~4,
GPT-4o), and the retest's Claude Sonnet~4.6 are objects of study.
Claude models (Anthropic)~\cite{anthropic2026claude} assisted with
initial checklist items, fault records, and ontology passages
(Section~\ref{sec:design}), and analysis code
(Section~\ref{sec:validation}). Claude and Codex assisted with
drafting and revising passages in Sections I--VI. The authors
conducted the research and evaluation, reviewed the instrument drafts
and revised the manuscript, and take responsibility for the reported
work. A Claude agent session designed, executed, and
analyzed the retest (\S\ref{sec:v3}) under author supervision.
Codex (OpenAI) ran the GPT-4o execution check, integrated the results,
and checked the revised manuscript build; a Codex ScholarPeer review
also informed the camera-ready revision. AI
agents (Claude Opus~5.5) performed the source audit; the first
author reviewed each flagged substance-error row, locating sources
with an AI search tool (Claude Opus~5.5) and deciding each after
reading them. Three AI reviewer agents on Claude Opus (ScholarPeer,
PaperViz, and a style-audit agent) commented on drafts. The
\S\ref{sec:kappa} primary adjudicator used no
AI tools; the first author transcribed the answers.


\begin{thebibliography}{10}
\providecommand{\url}[1]{#1}
\csname url@samestyle\endcsname
\providecommand{\newblock}{\relax}
\providecommand{\bibinfo}[2]{#2}
\providecommand{\BIBentrySTDinterwordspacing}{\spaceskip=0pt\relax}
\providecommand{\BIBentryALTinterwordstretchfactor}{4}
\providecommand{\BIBentryALTinterwordspacing}{\spaceskip=\fontdimen2\font plus
\BIBentryALTinterwordstretchfactor\fontdimen3\font minus
  \fontdimen4\font\relax}
\providecommand{\BIBforeignlanguage}[2]{{%
\expandafter\ifx\csname l@#1\endcsname\relax
\typeout{** WARNING: IEEEtran.bst: No hyphenation pattern has been}%
\typeout{** loaded for the language `#1'. Using the pattern for}%
\typeout{** the default language instead.}%
\else
\language=\csname l@#1\endcsname
\fi
#2}}
\providecommand{\BIBdecl}{\relax}
\BIBdecl

\bibitem{liu2023agentbench}
X.~Liu, H.~Yu, H.~Zhang, Y.~Xu, X.~Lei, H.~Lai, Y.~Gu, H.~Ding, K.~Men,
  K.~Yang, S.~Zhang, X.~Deng, A.~Zeng, Z.~Du, C.~Zhang, S.~Shen, T.~Zhang,
  Y.~Su, H.~Sun, M.~Huang, Y.~Dong, and J.~Tang, ``{AgentBench}: Evaluating
  {LLMs} as agents,'' in \emph{Proc. Int. Conf. Learning Representations
  (ICLR)}, 2024, pp. 52\,989--53\,046.

\bibitem{yao2024taubench}
S.~Yao, N.~Shinn, P.~Razavi, and K.~Narasimhan, ``$\tau$-bench: A benchmark for
  tool-agent-user interaction in real-world domains,'' in \emph{Proc. Int.
  Conf. Learning Representations (ICLR)}, 2025, pp. 9965--10\,017.

\bibitem{nist2023rmf}
{National Institute of Standards and Technology}, ``Artificial intelligence
  risk management framework ({AI} {RMF} 1.0),'' NIST AI 100-1, U.S. Department
  of Commerce, 2023.

\bibitem{vnailaw2025}
{National Assembly of Vietnam}, ``Law on {Artificial Intelligence}, {Law
  No.~134/2025/QH15},'' Passed December 10, 2025; in force March 1, 2026, 2025.

\bibitem{jia2011mutation}
Y.~Jia and M.~Harman, ``An analysis and survey of the development of mutation
  testing,'' \emph{IEEE Transactions on Software Engineering}, vol.~37, no.~5,
  pp. 649--678, 2011.

\bibitem{zheng2023judging}
L.~Zheng, W.-L. Chiang, Y.~Sheng, S.~Zhuang, Z.~Wu, Y.~Zhuang, Z.~Lin, Z.~Li,
  D.~Li, E.~P. Xing, H.~Zhang, J.~E. Gonzalez, and I.~Stoica, ``Judging
  {LLM}-as-a-judge with {MT-Bench} and {Chatbot Arena},'' in \emph{Advances in
  Neural Information Processing Systems 36 (NeurIPS), Datasets and Benchmarks
  Track}, 2023, pp. 46\,595--46\,623.

\bibitem{andriushchenko2024agentharm}
M.~Andriushchenko, A.~Souly, M.~Dziemian, D.~Duenas, M.~Lin, J.~Wang,
  D.~Hendrycks, A.~Zou, Z.~Kolter, M.~Fredrikson, Y.~Gal, and X.~Davies,
  ``{AgentHarm}: A benchmark for measuring harmfulness of {LLM} agents,'' in
  \emph{Proc. Int. Conf. Learning Representations (ICLR)}, 2025, pp.
  79\,185--79\,220.

\bibitem{guha2023legalbench}
N.~Guha, J.~Nyarko, D.~E. Ho, C.~R{\'e}, A.~Chilton, A.~Narayana,
  A.~Chohlas-Wood, A.~Peters, B.~Waldon, D.~N. Rockmore \emph{et~al.},
  ``{LegalBench}: A collaboratively built benchmark for measuring legal
  reasoning in large language models,'' in \emph{Advances in Neural Information
  Processing Systems 36 (NeurIPS), Datasets and Benchmarks Track}, 2023, pp.
  44\,123--44\,279.

\bibitem{zeng2025airbench}
Y.~Zeng, Y.~Yang, A.~Zhou, J.~Tan, Y.~Tu, Y.~Mai, K.~Klyman, M.~Pan, R.~Jia,
  D.~Song, P.~Liang, and B.~Li, ``{AIR-Bench} 2024: A safety benchmark based on
  regulation and policies specified risk categories,'' in \emph{Proc. Int.
  Conf. Learning Representations (ICLR)}, 2025, pp. 63\,997--64\,031.

\bibitem{harman2025ach}
M.~Harman, J.~Ritchey, I.~Harper, S.~Sengupta, K.~Mao, A.~Gulati, C.~Foster,
  and H.~Robert, ``Mutation-guided {LLM}-based test generation at {Meta},'' in
  \emph{Proc. ACM Int. Conf. Foundations of Software Engineering, Industry
  Track}, 2025, pp. 180--191.

\bibitem{ruan2024toolemu}
Y.~Ruan, H.~Dong, A.~Wang, S.~Pitis, Y.~Zhou, J.~Ba, Y.~Dubois,
  C.~J. Maddison, and T.~Hashimoto, ``Identifying the risks of {LM} agents
  with an {LM}-emulated sandbox,'' in \emph{Proc. Int. Conf. Learning
  Representations (ICLR)}, 2024.

\bibitem{borjigin2025goldmine}
A.~Borjigin, C.~He, C.~C. Lee, W.~Zhou, and C.~Y. Song, ``{AI} agent
  architecture for decentralized trading of alternative assets,'' in
  \emph{Proc. IEEE Int. Conf. Recent Advances in Systems Science and
  Engineering (RASSE)}, 2025, pp. 1--8.

\bibitem{chen2025shieldagent}
Z.~Chen, M.~Kang, and B.~Li, ``{ShieldAgent}: Shielding agents via verifiable
  safety policy reasoning,'' in \emph{Proc. Int. Conf. Machine Learning
  (ICML)}, vol.~267, 2025, pp. 8313--8344.

\bibitem{wang2026agentspec}
H.~Wang, C.~M. Poskitt, and J.~Sun, ``{AgentSpec}: Customizable runtime
  enforcement for safe and reliable {LLM} agents,'' in \emph{Proc. IEEE/ACM
  Int. Conf. Software Engineering (ICSE)}, 2026, pp. 2938--2950.

\bibitem{luong2026neurosymbolic}
T.~T. Luong and A.~Sanyal, ``Ontology-constrained neural reasoning in
  enterprise agentic systems: A neurosymbolic architecture for domain-grounded
  {AI} agents,'' \emph{arXiv preprint arXiv:2604.00555}, 2026.

\bibitem{vnamllaw2022}
{National Assembly of Vietnam}, ``{Law on Anti-Money Laundering},
  {No.~14/2022/QH15},'' 2022.

\bibitem{regatta2026archive}
T.~T. Luong and A.~Sanyal, ``{REGATTA ICoDSE} research artifact,''
  ver. 1.1.0, Zenodo, 2026, doi: 10.5281/zenodo.23006005.

\bibitem{regatta2026protocol}
T.~T. Luong and A.~Sanyal, ``REGATTA retest preregistration:
  Post-analysis Addendum 3,'' Sep. 28, 2026, accompanying
  supplementary protocol record.

\bibitem{anthropic2026claude}
{Anthropic}, ``Claude,'' product documentation. Accessed: Oct. 2, 2026.
  [Online]. Available: \url{https://claude.com/product/overview}

\end{thebibliography}
\end{document}